\documentclass[letterpaper,10pt,aps,reprint,prd,nofootinbib,noshowkeys,longbibliography,floatfix,superscriptaddress,preprintnumbers]{revtex4-1}
\usepackage{hyperref} 
\usepackage{amsmath}
\usepackage{amsfonts}
\usepackage{amsthm}
\usepackage{amssymb}
\usepackage{graphicx}
\usepackage{txfonts}
\usepackage{color}
\usepackage[sort&compress]{natbib}

\graphicspath{{./}{./pics/}}

\newcommand{\mr}[1]{\mathrm{#1}}

\newcommand{\ez}{\ensuremath{\epsilon_0}}

\begin{document}

\title{Observing Thermal Schwinger Pair Production}
\date{December 12, 2018}

\author{Oliver Gould}
\email{oliver.gould@helsinki.fi}
\affiliation{Department of Physics, Imperial College London, SW7 2AZ, UK}
\affiliation{Helsinki Institute of Physics, University of Helsinki, FI-00014, Finland}
\preprint{HIP-2018-31/TH}
\author{Stuart Mangles}
\email{stuart.mangles@imperial.ac.uk}
\author{Arttu Rajantie}
\email{a.rajantie@imperial.ac.uk}
\preprint{IMPERIAL-TP-2018-AR-2}
\author{Steven Rose}
\email{s.rose@imperial.ac.uk}
\author{Cheng Xie}
\email{cxie@live.com}
\affiliation{Department of Physics, Imperial College London, SW7 2AZ, UK}

\begin{abstract}
We study the possibility of observing Schwinger pair production enhanced by a thermal bath of photons. We consider the full range of temperatures and electric field intensities from pure Schwinger production to pure thermal production, and identify the most promising and interesting regimes. In particular we identify temperatures of $\sim 20~\mr{keV}/k_B$ and field intensities of $\sim 10^{23}~\mr{Wcm}^{-2}$ where pair production would be observable. In this case the thermal enhancement over the Schwinger rate is exponentially large and due to effects which are not visible at any finite order in the loop expansion. Pair production in this regime can thus be described as more nonperturbative than the usual Schwinger process, which appears at one loop. Unfortunately, such high temperatures appear to be out of reach of foreseeable technologies, though nonthermal photon distributions with comparable energy-densities are possible. We suggest the possibility that similar nonperturbative enhancements may extend out of equilibrium and propose an experimental scheme to test this.
\end{abstract}

\maketitle

\section{Introduction}

Schwinger long ago predicted that a strong, constant electric field will create electron-positron pairs \cite{schwinger1951gauge}. This has, however, never been experimentally observed as the electric field strengths required are several orders of magnitude larger than has ever been achieved in the laboratory \cite{danson2015petawatt,Turcu:2016dxm}. The rate of pair production becomes large as the electric field strength approaches an appreciable fraction of the Schwinger critical field, $E_c= m_e^2c^3/ e \hbar$, corresponding to an electric field intensity $I_c=\tfrac{1}{2}\ez c E_c^2 \approx 2\times 10^{29}~\mr{Wcm}^{-2}$. It has also long ago been predicted that one can create electron-positron pairs from a thermal bath of photons \cite{weaver1976reaction}, via the two photon Breit-Wheeler process, $\gamma\gamma\to e^+e^-$ \cite{breit1934collision}. This too has never been observed due to the unattainably high temperatures required (although an experiment has been proposed that uses a quasi-thermal radiation field for one of the two photons \cite{pike2014photon}). In this case, the rate of pair production becomes large when $k_B T$ becomes an appreciable fraction of the rest mass energy of an electron and positron, $2m_e c^2\sim 1~\mr{MeV}$.

One can, however, combine these two ingredients. Starting from an initial state containing a thermal bath of photons at a high temperature and adding a strong, constant electric field, one finds the rate of pair production is significantly faster than either the Schwinger process or the purely thermal process alone. The nature of the process depends on the relative magnitudes of the electric field strength and the temperature. At low temperatures, when $E/E_c\gg k_B T/2m_ec^2$, the process is essentially Schwinger pair production, in which virtual electron-positron pairs tunnel quantum mechanically through their energy barrier to existence. In the opposite limit, $E/E_c\ll k_B T/2m_ec^2$, at high temperatures, virtual electron-positron pairs are given sufficient energy from the thermal bath to transition over the barrier classically. At intermediate temperatures the process can be described as thermally enhanced quantum tunnelling, whereby the virtual electron-positron pairs tunnel from an excited state at nonzero separation.

Here we consider the viability of observing the thermal Schwinger process. If this were realised, it would be the first observation of semiclassical pair production in quantum electrodynamics (QED), a class of nonperturbative phenomena with applications in many branches of physics, including astrophysics, cosmology, heavy ion collisions, and plasma physics (see for example \cite{Ruffini:2009hg}). It would also open up the controlled study of semiclassical pair production in general, a very basic process in quantum field theory and one that has proved elusive experimentally.

To put our work in context, we note that several other mechanisms have been proposed to lower the intensities required for Schwinger pair production, by including high frequency fields \cite{brezin1970pair,Ringwald:2001ib,Popov:2001ak,Piazza:2004sv,dunne2009catalysis,Otto:2015gla,Panferov:2015yda,Torgrimsson:2016ant,Kohlfurst:2012rb,Jansen:2013dea}, or the Coulomb fields of highly charged nuclei \cite{Muller:2003zzd,PhysRevA.73.053409,DiPiazza:2009py,FillionGourdeau:2012zs}. Further, it has been pointed out that once an initial seed pair has been produced, a cascade of pair production will follow for sufficiently strong field intensities \cite{Bell:2008zzb,PhysRevLett.105.080402,Nerush:2010fe}. This may dramatically amplify any signal of Schwinger pair production. On the experimental side of strong-field QED, there has been much progress on a variety of fronts \cite{burke1997positron,mackenroth2011nonlinear,thomas2012strong,blackburn2014quantum,cole2018experimental,poder2018experimental}, and the next generation of high intensity lasers offer exciting possibilities to discover and investigate qualitatively new phenomena in QED \cite{DiPiazza:2011tq,Turcu:2016dxm}.

In the regime we consider here, there is an additional significance, in that the formula for the rate of pair production is a rare example of an all-orders and all-loop result in QED \cite{gould2017thermal,Gould:2018ovk}. Perturbative estimates of the rate of pair production are orders of magnitude slower, so could be distinguished experimentally. Hence, one can probe QED beyond the perturbative loop expansion, and could decide experimentally on the validity of conjectured all-order behaviours of QED \cite{cvitanovic1977asymptotic,affleck1981pair,lebedev1984virial,huet2017asymptotic,gould2017thermal,Gould:2018ovk}. Further, an experimental verification of the formulae would directly carry over to strongly-coupled physics, and hence provide an important validation of principle in the search for magnetic monopoles \cite{acharya2016search,gould2017magnetic}. As such, an experimental search for the thermal Schwinger process is well motivated even independently of its connection to the pure Schwinger process.

\section{Theoretical results}
At zero temperature, and for electric field strengths somewhat below $E_c$, the rate of electron-positron pair production per unit volume in a constant electric field is given by Schwinger's result \cite{schwinger1951gauge},
\begin{equation}
 \Gamma_{\mr{Schwinger}}(E)\approx \frac{(eE)^2}{4\pi^3 c \hbar^2}\mr{e}^{-\frac{\pi m_e^2 c^3}{e  E \hbar}}. \label{eq:rate_schwinger}
\end{equation}
Loop corrections to this formula have been computed at two loops \cite{ritus1975lagrange,ritus1977connection,lebedev1984virial,ritus1998effective}, (see also \cite{Gies:2016yaa}) and have been resummed to all loops (within the quenched approximation) \cite{affleck1981pair,lebedev1984virial}. However, the loop corrections are small, and only give an $O(1\%)$ enhancement over Schwinger's one-loop result. At leading loop order, there are also corrections which are exponentially subdominant for small $E/E_c$ \cite{schwinger1951gauge}.

If one adds a thermal bath of photons, the rate is enhanced. Note that in this analysis it is crucial that there are very few charged particles in the initial thermal state as their presence would Debye screen the electric field. The Debye screening length should be longer than the scales relevant for pair production (which we give in Section \ref{sec:schematic}). Hence, we assume $k_B T\ll m_e c^2$.

Depending on the relative magnitudes of $E$ and $T$ one finds three different regimes. In the lowest temperature regime, the energy of the thermal bath is less than the energy that an electron-positron pair would gain when accelerated by the electric field over their Compton wavelength. That is, for temperatures lower than
\begin{equation}
    T_{CW}:=e E\hbar/(2m_e c k_B). \label{eq:lowT}
\end{equation}
In this regime, the thermal bath is negligible and electron-positron pairs are produced by quantum tunnelling from virtuality in vacuum, to reality. Above $T_{CW}$ the thermal bath excites virtual electron-positron pairs significantly above their ground state. Pair production then takes the form of quantum tunnelling from an excited state. At higher temperatures still, virtual electron-positron pairs acquire sufficient energy from thermal fluctuations to go over the energy barrier classically. This process dominates over quantum tunnelling for temperatures greater than
\begin{equation}
    T_{WS}:=\left(4 e E^3\ez\hbar^4/\pi^3 m_e^2 k_B^4\right)^{1/4}
\end{equation}
This temperature can be understood as that when the lowest thermal (Matsubara) frequency $2\pi k_BT/\hbar$ is equal to the exponential decay rate of the unstable electron-positron transition state, $2\sqrt{2}(eE^3\ez/m_e^2)^{1/4}$.\footnote{The labels $C$, $W$ and $S$ follow the notation of Ref.~\cite{gould2017thermal}. They stand for \emph{circular}, \emph{wavy} and \emph{straight} (or \emph{sphaleron}) respectively, and refer to the shape of the instanton describing the processes at low, intermediate and high temperatures.}

\begin{figure}
  \includegraphics[width=1.0\columnwidth]{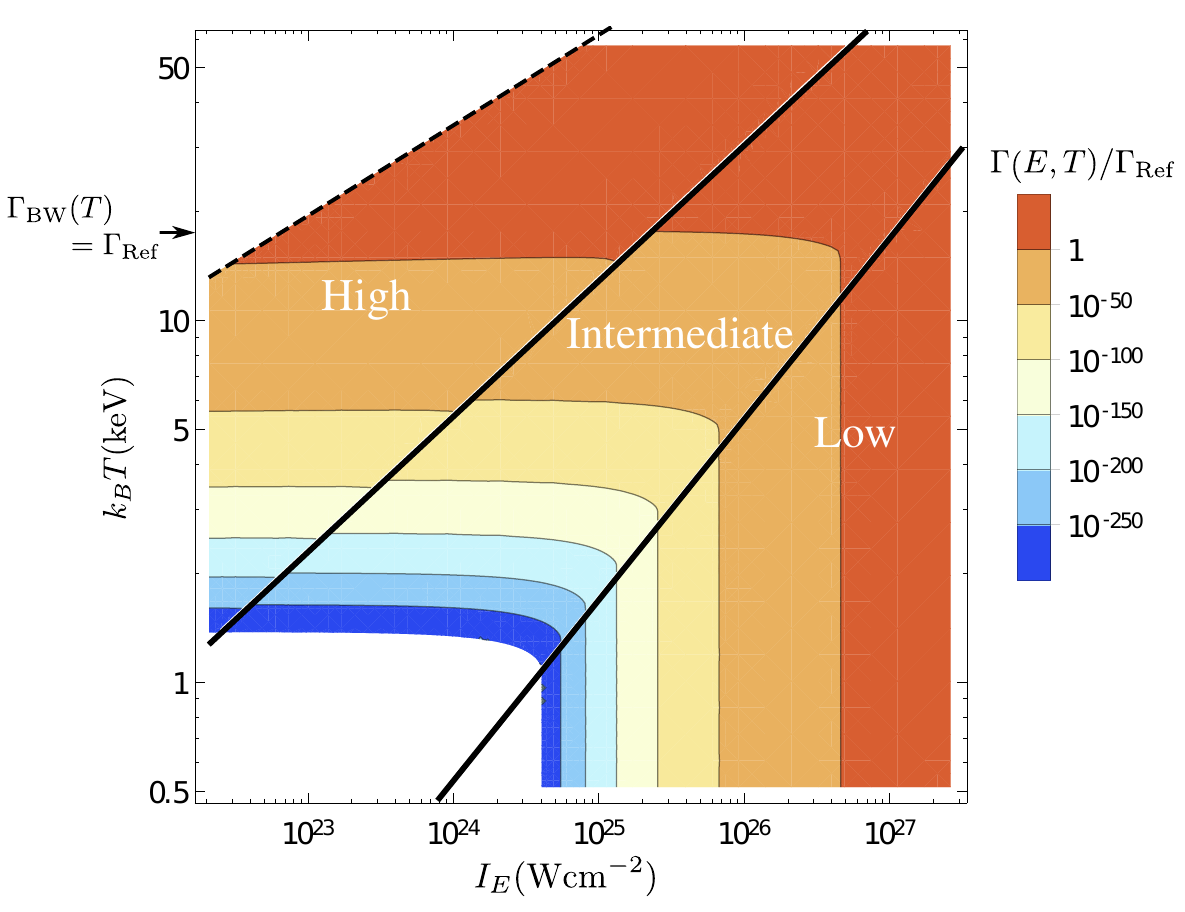}
  \caption[Thermal Schwinger Rate]{Approximate rates of pair production for low, medium and high temperatures. The coloured region is where the approximations leading to Eqs. \eqref{eq:rate_schwinger}, \eqref{eq:rate_w} and \eqref{eq:rate_sphaleron} are well satisfied. All rates are normalised by $\Gamma_{\mr{Ref}}=0.1~\muup\mr{m}^{-3}\muup\mr{s}^{-1}$. For intermediate temperatures we use only the leading two contributions in the fine structure constant whereas in the high temperature regime all orders are included. This leads to the apparent discontinuity which will be smoothed out by contributions at higher orders.}
  \label{fig:allRatesApprox}
\end{figure}

\begin{figure}
  \includegraphics[width=0.7\columnwidth]{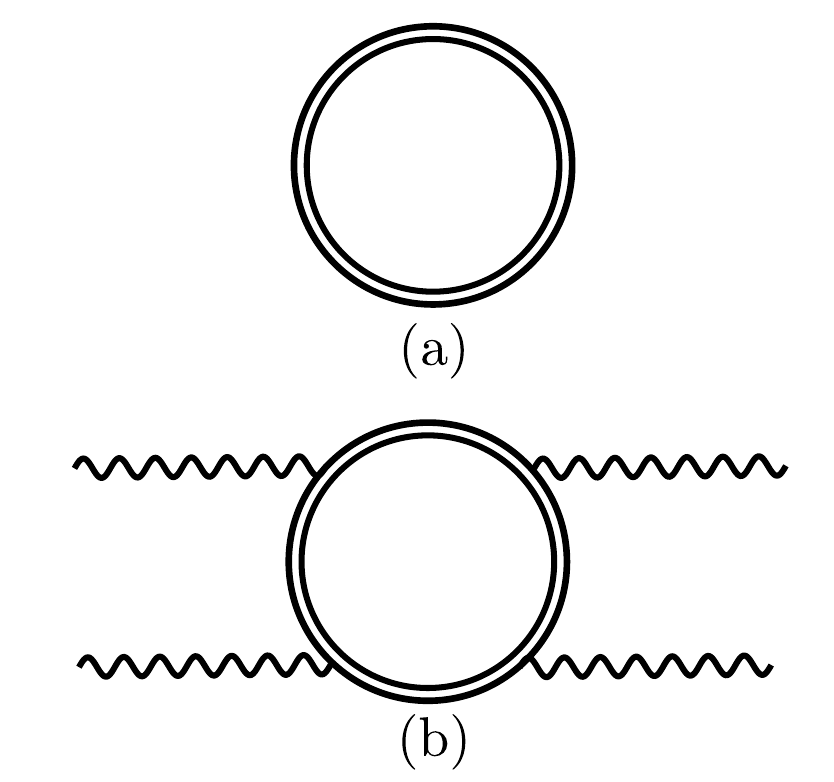}
  \caption{Feynman diagrams which dominate the rate of thermal Schwinger pair production in the (a) low and (b) intermediate temperature regimes. Double lines denote the electron propagator including the effect of the external electric field to all orders and the wiggly lines denote photons from the thermal bath. In the high temperature regime, the rate is not dominated by a single such Feynman diagram but infinitely many diagrams contribute to the leading approximation to the rate.}
  \label{fig:feyn}
\end{figure}

\emph{Low temperatures,} $T<T_{CW}$:
For very low temperatures, $T\ll T_{CW}$, corrections to the Schwinger rate can be considered perturbatively. The dominant contribution to the rate in this regime is thus given by Fig.~\ref{fig:feyn}~(a), just as at $T=0$. The leading thermal corrections arise at two loops and are \cite{gies2000qed}
\begin{equation}
    \Gamma_C=\Gamma_{\mr{Schwinger}}\left[1+\frac{\pi^4 e^2}{1440\ez\hbar c} \left(\frac{T}{T_{CW}}\right)^4+\dots\right].
\end{equation}
As $T$ increases towards $T_{CW}$, higher order perturbative corrections become important. The exponential enhancement due to these corrections has been computed at leading \cite{monin2010photon} and next-to-leading \cite{gould2017thermal} order. The corrections to the exponent of the rate are small if $E\lesssim E_c$, except if one is very close to $T_{CW}$ in which case even the exponent is unknown. Where one can trust the calculations, the thermal enhancement at low temperatures is less that 1\%.

\emph{Intermediate temperatures,} $T_{CW}<T<T_{WS}$:
At $T_{CW}$, the rate goes through a sharp transition. In this intermediate temperature range, the rate of pair production is significantly (exponentially) enhanced by the presence of the thermal bath. For intermediate temperatures, the exponent of the rate is given by \cite{selivanov1986tunneling,ivlev1987tunneling,brown2015schwinger,gould2017thermal,Draper:2018lyw}
\begin{align}
  &\Gamma_{W}(E,T)\sim \nonumber \\
  &\exp\left[-2\frac{m_e^2c^3}{eE\hbar}\left( \arcsin\left(\frac{T_{CW}}{T}\right) +  \frac{T_{CW}}{T} \sqrt{1 - \frac{T_{CW}^2}{T^2}}\right)\right], \label{eq:rate_w}
\end{align}
though there has been some dispute on this \cite{medina2015schwinger,Korwar:2018euc}. The precise formula, including the prefactor of the exponential has recently been worked out in Ref.~\cite{Torgrimsson:2019sjn}. There it was also shown that the dominant contribution to the rate in this regime is given by Fig.~\ref{fig:feyn}~(b).

\emph{High temperatures,} $T_{WS}<T$:
At $T_{WS}$ there is again a sharp transition in the rate. In the high temperature regime, the pair production process can be seen as a thermal process enhanced by the presence of the electric field. Unlike the lower temperature regimes, the rate is not dominated by the diagrams in Fig.~\ref{fig:feyn}; infinitely more such diagrams contribute to the leading approximation to the rate, leading to a significant nonperturbative enhancement. Some of the present authors have recently calculated the rate \cite{gould2017thermal,Gould:2018ovk}. It is given by
\begin{equation}
 \Gamma_{S}(E,T)\approx \frac{4 k_B T_{WS}\left(m_e k_B T\right)^{3/2}\mr{e}^{-\frac{2m_e c^2}{k_B T}+\frac{\sqrt{e^3E/\pi\ez}}{k_B T}}}{(4\pi)^{3/2}\hbar^4\sin \left(\frac{\pi T_{WS} }{T }\right)\sinh^2\left(\frac{\pi T_{WS}}{\sqrt{2} T }\right)} .\label{eq:rate_sphaleron}
\end{equation}
In deriving this expression we assumed the following three strong inequalities: $e E\hbar/m_e^2c^3\ll 1$, $k_BT/m_e c^2\ll 1$ and $e (k_B T)^2 /E c^2 \ez \hbar^2\ll 1$. Note that these imply the calculation is only valid when the rate of pair production is not too fast. For temperatures much larger than $T_{WS}$, this equation simplifies to 
\begin{align}
\Gamma_S(E,T) & \approx \frac{m_e^{3/2} T^2 \left(k_B T\right)^{5/2}}{\pi^{9/2}\hbar^4 T_{WS}^2}  \mr{e}^{-\frac{2m_e c^2}{k_B T}+\frac{\sqrt{e^3E/\pi\ez}}{k_B T}} \nonumber \\
&\approx \frac{ m_e^{5/2} (k_B T)^{9/2}}{ 2 \pi ^3 \hbar^{6} \sqrt{ \ez e E^{3}} } e^{-\frac{2 m_e c^2}{k_BT}+\frac{\sqrt{e^3E/\pi\ez}}{k_B T}}.\label{eq:rate_sphaleron_high}
\end{align}
As one can see from this expression, surprisingly the rate is higher for weaker fields. This behaviour cannot be extrapolated to arbitrarily weak fields as the validity of our approximations breaks down for $E\lesssim e (k_B T)^2 /c^2 \ez \hbar^2$.

For a thermal bath of photons in zero electric field, electron-positron pairs can be produced by two photon fusion (the Breit-Wheeler process). Integrating the cross section for this process over the photon thermal distribution one finds \cite{weaver1976reaction}\footnote{Note that there are several typos in Ref.~\cite{weaver1976reaction}. We have repeated the calculation for low temperatures, both fully numerically and in the nonrelativistic approximation, finding agreement with Eq. \eqref{eq:breit_wheeler}.}
\begin{align}
 \Gamma_{BW}(T)&\approx 2 \int \frac{d^3 p}{(2\pi )^3}\frac{d^3 q}{(2\pi )^3} \frac{p^\mu q_\mu}{p^0 q^0}\theta\left((p^\mu+q^\mu)^2-4m^2\right)\nonumber \\
 & \qquad\qquad\qquad\qquad f(p^0)f(q^0)\sigma_{BW}\left((p^\mu+q^\mu)^2\right)\nonumber \\
 &\approx \frac{e^4 m_e (k_B T)^3}{2(2\pi)^4 c^3 \ez^2 \hbar^6}\mr{e}^{-\frac{2m_e c^2}{k_B T}},\label{eq:breit_wheeler}
\end{align}
where $f(E)=1/(\mr{e}^{E/k_BT}-1)$ and $\sigma_{BW}(s)$ is the Breit-Wheeler cross section \cite{breit1934collision}. This expression is valid for low temperatures, $k_B T/m_e c^2 \ll 1$, up to about $O(100~\mr{keV}/k_B)$ with an accuracy of a few percent. In the absence of an intense electric field, higher order perturbative effects involving more photons are expected to be subdominant if $e k_B T/\sqrt{c \hbar \ez} \ll m_ec^2$, or $k_BT\ll 3 m_e c^2$ \cite{king2012pair}.

We note that the addition of a constant electric field does not enhance the perturbative Breit-Wheeler process as a constant field consists of zero energy photons. Beyond this idealised limit, corrections to the Breit-Wheeler rate due to the presence of an additional source of photons with wavelength $\lambda$ are suppressed by $\exp(-\lambda m^2 c^3/(2\pi \hbar k_B T))$. For a typical laser source with $\lambda = 0.8~\muup \mr{m}$ and a thermal bath of temperature $T=20~\mr{keV}/k_B$, say, this correction is completely negligible $\sim 10^{-10^6}$.

The Breit-Wheeler process should not be thought of as a competing process to thermal Schwinger pair production. Fig.~\ref{fig:feyn}~(b) shows the dominant Feynman diagram for thermal Schwinger pair production in the intermediate temperature regime. Applying the Optical Theorem, one can see that a unitarity cut of diagram (b) gives the Breit-Wheeler process~\cite{Torgrimsson:2019sjn}, except that the effect of the electric field has been accounted for to all orders, giving a nonperturbative enhancement over the pure Breit-Wheeler process. In fact, Eq.~\eqref{eq:rate_w} reduces to Eq.~\eqref{eq:breit_wheeler} for temperatures $T\gg T_{CW}$.

However, in the high temperature regime, $T>T_{WS}$, the diagrams of Fig.~\ref{fig:feyn} cease to dominate the rate of pair production and there is a further nonperturbative enhancement that cannot easily be understood diagramatically. As can be seen from Eqs.~\eqref{eq:rate_sphaleron} and \eqref{eq:rate_sphaleron_high}, the dependence of the rate, $\Gamma_S$, on the fine-structure constant is non-analytic even after one absorbs one power of $e$ into the electric field.

For completeness, we note that the addition of a single electromagnetic plane wave to a thermal bath of photons also leads to nonperturbatively enhanced electron-positron pair production. This is true even in the long-wavelength limit, $\lambda m c/\hbar \to \infty$, showing the collective, nonperturbative nature of the phenomenon. In this case, the Breit-Wheeler rate is additively enhanced by \cite{king2012pair},
\begin{equation}
\Gamma_{\mr{Plane}} \approx \frac{3^{3/4} e^2 (k_B T)^2 m^2 }{16 \pi ^{5/2} \ez \hbar^5} \left(\frac{e E \hbar k_B T}{
   m^3 c^5}\right)^{1/4} 
   \mr{e}^{-\sqrt{\frac{16c^5 m^3}{3 e E \hbar k_B T }}}. \label{eqn:rate_king} 
\end{equation}
This result is valid for $\sqrt{k_B T E/(m c^2 E_c)}\ll 1$. The crucial difference from that of a constant electric field is that the electromagnetic invariant $E^2-c^2B^2$ of a plane wave vanishes. As we will note later, Eq. \eqref{eqn:rate_king} is orders of magnitude smaller than the thermal Schwinger rate, showing that the absence of the magnetic field is crucial for pair production.

\section{Observability}

We would like to understand exactly how high the temperatures and how strong the electric fields need to be to get a measurable rate of pair production. To answer that, we will make a simple comparison with the experiment of Ref.~\cite{burke1997positron}, which was the first experiment to observe the (multi-photon) Breit-Wheeler process. They observed $106\pm14$ positrons produced in this way, from a total spacetime interaction volume of order $10^{-21}~\mr{m}^3 \mr{s}$ (when integrated over all laser shots). Hence we take as our observable reference rate $\Gamma_{\mr{Ref}}=10^{23} ~\mr{m}^{-3}\mr{s}^{-1}=0.1~\muup\mr{m}^{-3}\muup\mr{s}^{-1}$, which is approximately Avogadro's number of positrons per metre cubed per second. One can therefore reasonably expect that a normalised rate greater than 1 will be required for the rate of pair production to be measurable. In Fig.~\ref{fig:allRatesApprox} we show the thermal Schwinger rate in all three regimes, normalised by this reference rate.

The almost perfectly vertical lines of constant rate in the low temperature regime reflect that, in this regime, the thermal enhancements are small. As such, this regime offers no advantages over pure Schwinger pair production for experimentally observing pair production. On the other hand, in the intermediate and high temperature regimes, the thermal enhancements are very significant. Of these two regimes, observing pair production in the high temperature regime is easier, because the electric field intensities required are orders of magnitude smaller, while the temperatures required are very similar.

From Figure \ref{fig:allRatesApprox} one can see that temperatures around $O(20~\mr{keV}/k_B)$ or above are needed in order to produce an observable number of positrons. Perhaps the leading method of producing high temperature thermal photons is with a laser and cavity, or holhraum. The aim of achieving inertial confinement fusion (ICF) has been a powerful incentive in developing these technologies. Thermal distributions of $0.3~\mr{keV}/k_B$ have been achieved since 1990, though about $0.4~\mr{keV}/k_B$ is likely the upper limit of this approach \cite{lindl2004physics}. Unfortunately at these temperatures, the thermal enhancement of the Schwinger rate is negligible.

When ICF is achieved, the burning thermonuclear plasma leads to significantly higher energy densities. Charged particles in the plasma are expected to reach temperatures from $O(20~\mr{keV}/k_B)$ to $O(200~\mr{keV}/k_B)$, depending on the composition and size of the burning plasma \cite{tabak1996role,rose2013electron}. Burning deuterium (D) plasmas are expected to be hotter than burning deuterium-tritium (DT) plasmas, as the peak nuclear reaction rate is at higher energies for D-D nuclear reactions. For a fixed composition, larger plasmas reach higher temperatures.

As the plasma is not optically thick, the effective temperature of the photons is lower than that of the charged particles. For representative examples, of burning deuterium plasma with radii $r=120~\muup\mr{m}$ and $r=150~\muup\mr{m}$, one finds that the photon energy density is equal to that of a Planck distribution with two degrees of freedom at $T=22~\mr{keV}/k_B$ and $T=26~\mr{keV}/k_B$ respectively. The photon distribution can be calculated using the approach of Ref.~\cite{rose2013electron}. However, the result is further from equilibrium than that of the charged particles. For now, we will assume a thermal distribution of photons, though we will return to this point in Section \ref{sec:distributions}.

\section{An experimental scheme} \label{sec:schematic}

\begin{figure}
  \includegraphics[width=1.0\columnwidth]{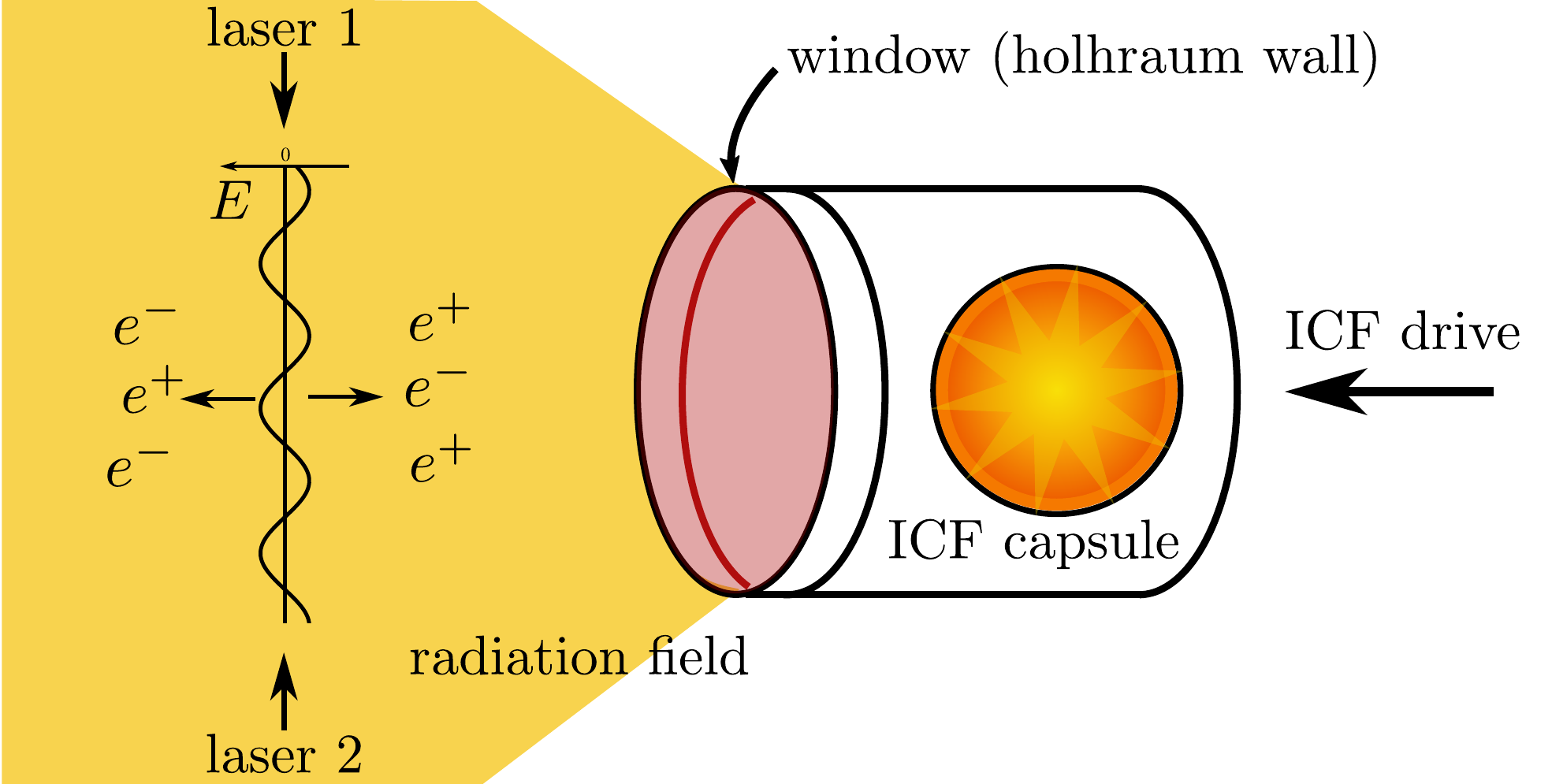}
  \caption{Schematic of the experimental set-up. Two counter propagating high energy beams are focused into an X-ray radiation field produced by a burning fusion.}
  \label{fig:schematic}
\end{figure}

So, in order to observe the thermal Schwinger process, we would propose combining two lasers with combined intensity $O(10^{23}~\mr{Wcm}^{-2})$ with a source of thermal photons with temperature $O(20~\mr{keV}/k_B)$. A possible schematic for such an experiment is shown in Figure \ref{fig:schematic}. The region of interest for our purposes is on the left-hand side, outside the ignited thermonuclear plasma. A window is needed to hold up the material expansion long enough to allow the high-intensity lasers to interact with the radiation from the ICF capsule. The wall of the hohlraum would in principle be able to act in this way whilst transmitting the majority of the radiation, though this would require specific design. As long as the distances from the nuclear plasma are small compared with its radius, the geometric reduction of the intensity will not be significant.

The electric field is provided by a high intensity laser, split into two counter-propagating beams. These are focused so that the magnetic fields cancel in the vicinity of a given point, whereas the electric fields reinforce. Assuming standard parameters for the laser, with wavelength $\lambda \sim 0.8~\muup m$, the field maxima of the two beams are expected to be approximately of size $O(\lambda^3)$ and of time extent $O(\lambda/c)$, with approximately 10-20 field maxima per shot, amounting to a possible pair production region of size $5\times 10^{-32}~\mr{m}^3\mr{s}$. The integrals of the rate over the interaction region can be carried out in the locally constant field approximation (see for example Refs.~\cite{galtsov1983macroscopic,Gavrilov:2016tuq}), within which the region around the field maxima will dominate the pair production. However, in what follows we simply multiply the rates by the approximate spacetime volume of the field maxima, which is sufficient to get the order of magnitude correct.

\begin{figure}
  \includegraphics[width=1.0\columnwidth]{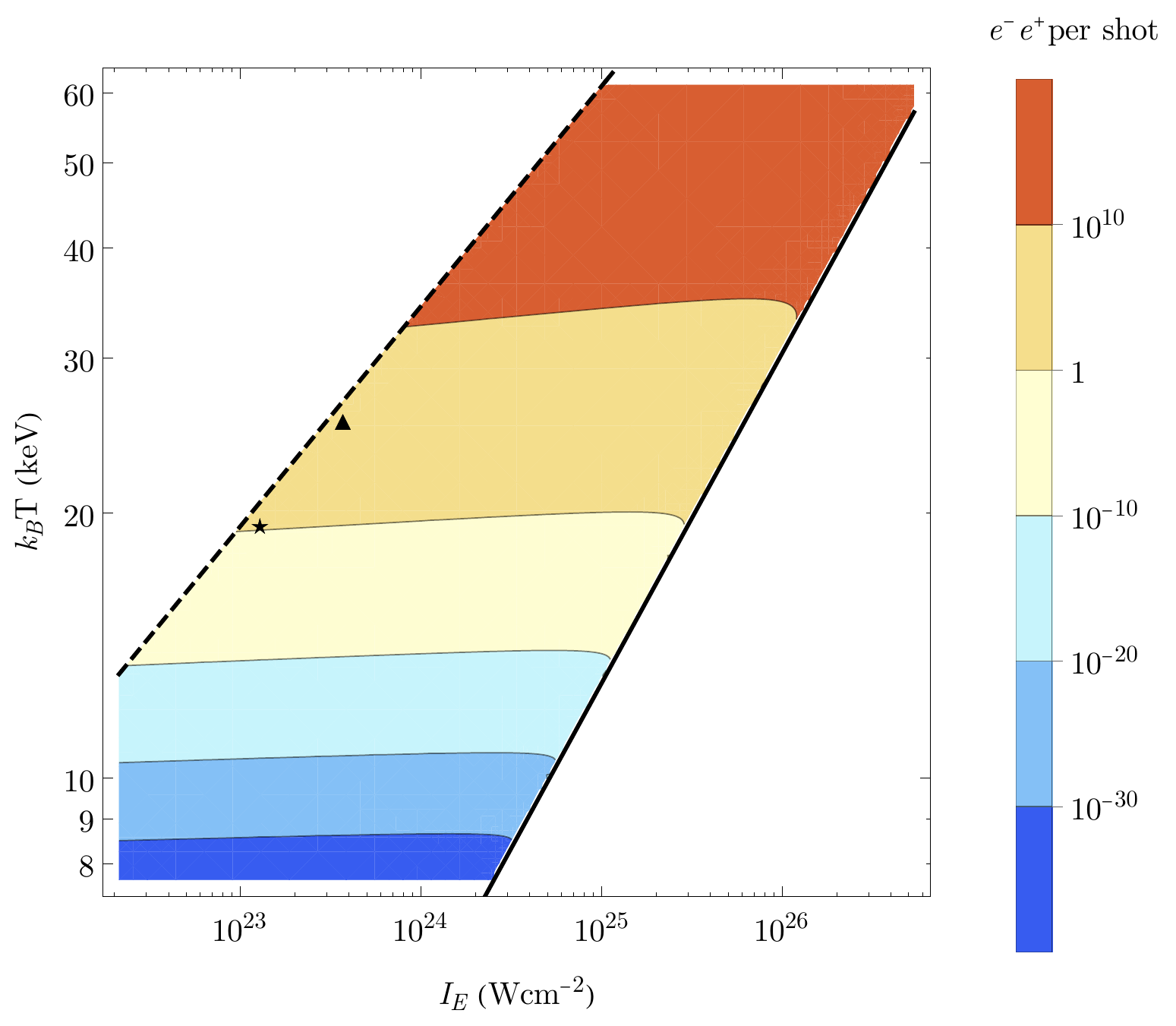}
  \caption[Thermal Schwinger rate at high temperature]{The number of electron-positron pairs produced per shot in the experiment proposed here. The coloured region is the high temperature region, and where the approximations leading to Eq. \eqref{eq:rate_sphaleron} are valid. The solid black line is the boundary between the intermediate and high temperature regions, defined by $T=T_{WS}$. The dashed black line is defined by $E=0.2e(k_BT)^2/\ez c^2 \hbar^2$ and the dotted black line by $T=0.2m_ec^2/k_B$. The star and diamond are the points referred to in Eqs. \eqref{eq:parameters} and \eqref{eq:parameters_triangle}.}
  \label{fig:sphaleron_rate}
\end{figure}

To achieve $\gtrsim 1$ electron-positron pair produced per shot, requires a rate $5\times 10^6$ times faster than $\Gamma_{\mr{Ref}}$ (see Fig.~\ref{fig:sphaleron_rate}). Assuming a thermal distribution of photons from the burning nuclear plasma, this could be achieved at
\begin{align}
 T_\star &\approx 20~\mr{keV}/k_B, \nonumber \\
 I_{E\star} &\approx 1.3\times 10^{23}~ \mr{W cm}^{-2}, \label{eq:parameters}
\end{align}
where $I_E$ refers to the combined intensity of the two beams. This parameter point is shown as a star in Fig.~\ref{fig:sphaleron_rate}. For comparison, we also consider a second point with a significantly higher production rate, at 
\begin{align}
 T_\blacktriangle &\approx 26~\mr{keV}/k_B, \nonumber \\
 I_{E\blacktriangle} &\approx 3.7\times 10^{23}~\mr{W cm}^{-2}. \label{eq:parameters_triangle}
\end{align}
For these two sets of parameters, the numbers of positrons produced per shot via the thermal Schwinger, Breit-Wheeler and pure Schwinger (without thermal enhancement) processes are given in Table \ref{table:positrons_per_shot}. We also include the nonperturbative enhancement to the number of positrons produced due to only one of the two laser beams, given by Eq. \eqref{eqn:rate_king}. 

Note that the pure Breit-Wheeler process can take place in a larger region than that of the Schwinger pair production, which is not accounted for in Table~\ref{table:positrons_per_shot}. In order to ensure that the thermal Schwinger process dominates, and taking into account its $O(10^6)$ times higher rate, the volume of the interaction region should be significantly less than about $10^7 \lambda^3\approx 5\times 10^{-3}\mr{mm}^3$. This can be achieved by modifying the diameter of the hohlraum window and by focusing the laser fairly close to the window. Further, the directionality of emitted electrons and positrons can help distinguish between production mechanisms, with the Breit-Wheeler process producing pairs more or less isotropically and the Schwinger process producing pairs along the electric field of the counterpropagating lasers.

It is encouraging that a relatively small increase in both radiation temperature and laser intensity
produces such a significant increase in the production rate. One can also see that the thermal Schwinger process has a huge nonperturbative enhancement. A simple perturbative estimate of the number of positrons produced by the Breit-Wheeler process underestimates the actual number by a factor of $10^6$. Such large enhancements are a generic feature of the thermal Schwinger process in the high temperature regime \cite{Gould:2018ovk}.

\begin{table}
\begin{center}
\begin{tabular}{ c c c c c } 
  \toprule
   & Thermal Schwinger & Breit-Wheeler & Eq. \eqref{eqn:rate_king} & Schwinger\\
  \colrule
  $\star$ & 3 & $3\times 10^{-6}$ & $10^{-167}$ & $10^{-1817}$\\
  $\blacktriangle$ & $1\times 10^6$ & 1 & $10^{-105}$ & $10^{-1062}$ \\
\end{tabular}
\caption{Numbers of positrons produced per shot via different mechanisms for the two sets of parameters given by Eqs. \eqref{eq:parameters} and \eqref{eq:parameters_triangle}. The column labelled Eq. \eqref{eqn:rate_king} is that for a thermal bath plus a single laser beam (rather than counter-propagating beams).}
\label{table:positrons_per_shot}
\end{center}
\end{table}

One can also compare the thermal Schwinger rate to that obtained in a thermal bath plus only a single high intensity laser beam (in which case the magnetic field does not cancel). In this case, the enhancement of the rate of pair production due to the high intensity laser is given by Eq.~\eqref{eqn:rate_king}. For the parameters of either Eqs. \eqref{eq:parameters} or \eqref{eq:parameters_triangle}, one finds that the enhancement is negligible and the rate of this process is smaller than the thermal Schwinger rate by a factor of $\sim 10^{−100}$ or more, as can be seen in Table \ref{table:positrons_per_shot}.

For the validity of the locally constant field approximation, it is important that the electric field, as well as the photon distribution from the plasma, are slowly varying on the time and length scales of the pair creation process, described by an instanton. The time scale of the instanton is $t_{\mr{inst}}\sim \hbar/k_B T$ and the length scale is $\sqrt{e/(4\pi \ez E)}$ \cite{gould2017thermal,Gould:2018ovk}. Using temperature and electric field strengths determined by Eq. \eqref{eq:parameters}, this amounts to $3\times 10^{-20}~s\approx 10^{-11}~\mr{m}/c$ and $5\times 10^{-12}~\mr{m}$ respectively. The smallest length scale on which the electric field varies is the wavelength of the laser. Assuming a laser with wavelength of $\lambda \sim 0.8~\muup m$, one can safely treat the electric field as constant. Further, one would expect the photon distribution from the plasma to vary on a length scale of order the size of the hohlraum window. This will likewise be much larger than the length scale of the instanton, $\sim 5\times 10^{-12}~\mr{m}$, and hence the locally constant field approximation is applicable.

In the region where the electric field and thermal photons collide, electrons and positrons will be produced with an approximately thermal spectrum of velocities and then accelerated in opposite directions antiparallel and parallel respectively to the electric field. Their thermal velocities are isotropic in the lab frame and are expected to be rather large, $\tfrac{1}{3}\bar{v^2}\sim k_B T/m_e \sim (0.2 c)^2$. The field then accelerates the particles over a distance $\lesssim \lambda$, giving them a highly relativistic velocity, $v\approx c$, parallel to the electric field and up to energies of order $eE\lambda \sim 1~\mr{GeV}$. Once produced, the electrons and positrons may be deflected in opposite directions with a magnet, after which their momenta can be measured by a calorimeter, as in Ref. \cite{burke1997positron}. If the combined intensity of the lasers is greater than around $10^{24}~\mr{Wcm}^{-1}$, a seed electron-positron pair produced by the thermal Schwinger process will induce a cascade of pair production, so amplifying any positive signal~\cite{Bell:2008zzb,PhysRevLett.105.080402,Nerush:2010fe}.

In the absence of charged particles, the thermal Schwinger process is the dominant mechanism of electron-positron pair production. However, if charged particles are not adequately shielded, other pair production processes are possible, such as the trident mechanism ($e^-Z\to e^- e^+e^- Z$) and the Bethe-Heitler process ($\gamma Z\to e^+e^- Z$). Another possibility is for non-linear Compton scattering of charged particles in the laser field, producing high energy photons which then take part in the Breit-Wheeler process. Debye screening by charged particles will also inhibit the thermal Schwinger process if the Debye length is not much longer than the length scale of the pair creation process, $\sqrt{e/(4\pi \ez E)}$. For the parameters of Eq. \eqref{eq:parameters}, one requires the density of charged particles to be much less than one per $\mr{pm}^3$. In the regime we have considered here, the purely thermal and the purely Schwinger pair production rates are orders of magnitude lower than the combination. Thus by performing null shots, with either only the burning plasma or only the high intensity laser, one can measure the presence of any backgrounds.

\section{Photon distributions}\label{sec:distributions}

\begin{figure}
  \includegraphics[width=1.0\columnwidth]{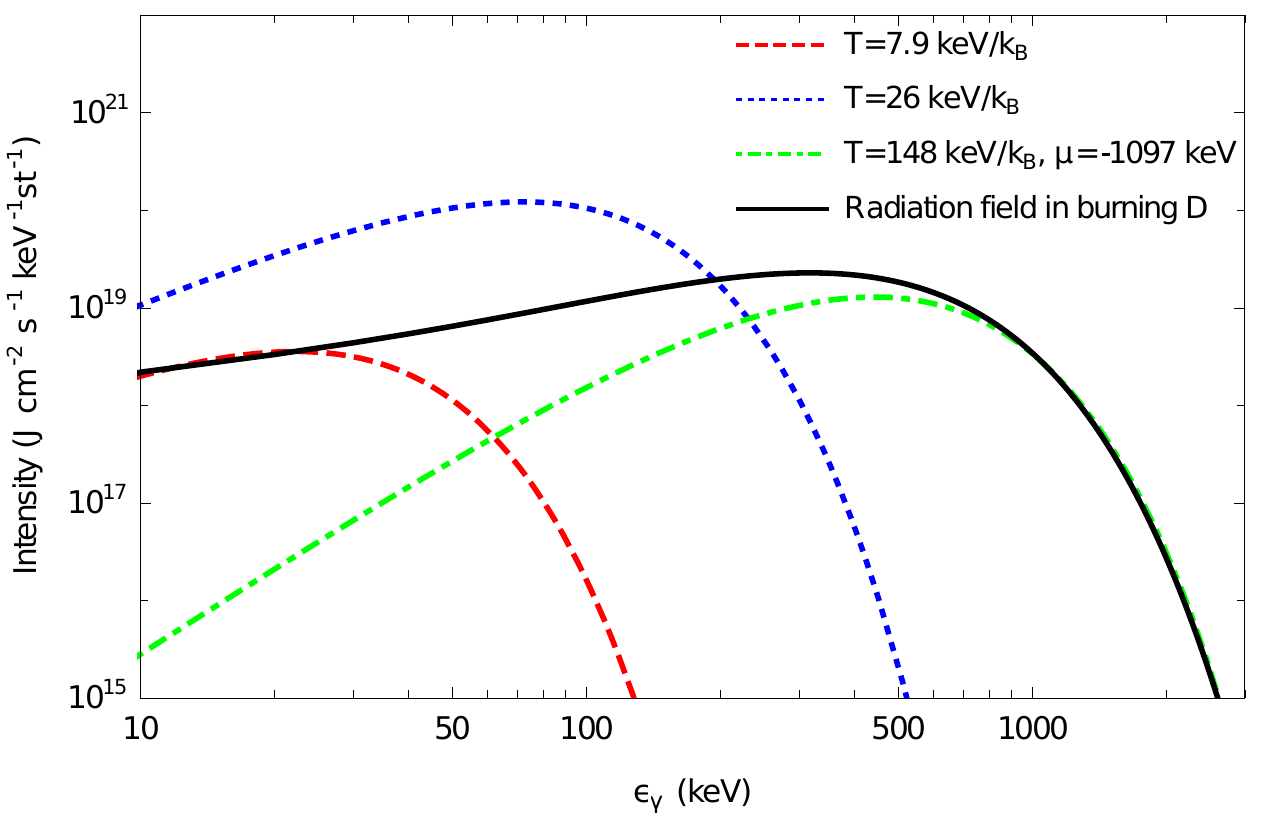}
  \caption[Photon Intensity]{Photon intensity in a D-burning target of radius $150~\muup\mr{m}$, along with various approximations to it. Note that a purely thermal distribution at $T=148~\mr{keV}/k_B$ would lie at much higher intensities.}
  \label{fig:intensity}
\end{figure}

Let us return to consider the distribution of photons in the burning plasma. This must be close to equilibrium for our approach to be valid. To investigate this we have solved the Boltzmann equation for the distribution of photons for a range of different plasma sizes and compositions. We have followed the method of Ref.~\cite{rose2013electron}, including the effect of Compton scattering. For our representative example, of a burning deuterium plasma of radius $r=150~\muup\mr{m}$, the photon intensity at the surface of this plasma is shown as the full black line in Fig.~\ref{fig:intensity}.

Equating the photon energy density to that of a thermal distribution, one finds that the effective temperature of the distribution is $26~\mr{keV}/k_B$. Doing the same for the photon number density, one instead finds a somewhat lower effective temperature of $16~\mr{keV}/k_B$, showing that the distribution is shifted to higher energies with respect to a thermal distribution. Plotting the photon intensity of a Planck distribution at $T=26~\mr{keV}/k_B$, the blue dotted line in Fig.~\ref{fig:intensity}, one can see the shift to higher energies.

The high energy tail of the distribution, above about $700~\mr{keV}$, is an exponential fall off and hence fits well a Boltzmann tail with an effective temperature of $T=148~\mr{keV}/k_B$, though scaled down by a normalisation, or equivalently a negative photon chemical potential\footnote{The photon chemical potential must be zero in equilibrium, but not necessarily out of equilibrium. In this context its presence is natural as photon number conserving processes dominate.}, $\mu=-1097~\mr{keV}$, plotted as the dot-dashed green line in Fig.~\ref{fig:intensity}. At the lowest energies, the distribution rises above this, and can be better described by a purely thermal distribution at a much lower temperature, $T=7.9~\mr{keV}/k_B$, plotted as the dashed red line in Fig.~\ref{fig:intensity}. At intermediate energies, the distribution is not well described by a Bose-Einstein distribution. Nevertheless, the overall shape of the distribution is qualitatively similar to a thermal distribution, being smooth and highly occupied with a power-like rise at low energies and an exponential decrease at high energies, though we have used four different effective temperatures to describe different aspects of it, ranging from $7.9~\mr{keV}/k_B$ to $148~\mr{keV}/k_B$.

In two counterpropagating laser beams with intensity given by Eq.~\eqref{eq:parameters} or \eqref{eq:parameters_triangle}, one finds that the intermediate temperature regime of thermal Schwinger pair production would be reached at temperatures above
\begin{align}
    T_{CW,\star} &=0.20~\mr{keV}/k_B, \nonumber \\
    T_{CW,\blacktriangle} &= 0.32~\mr{keV}/k_B, \label{eq:tcw} 
\end{align}
and the high temperature regime would be reached at temperatures above
\begin{align}
    T_{WS,\star} &=2.5~\mr{keV}/k_B, \nonumber \\ 
    T_{WS,\blacktriangle} &= 3.7~\mr{keV}/k_B. \label{eq:tws} 
\end{align}
All four effective temperatures we have used to describe the distribution of photons in the burning plasma are well above these temperatures. We thus expect the high temperature regime to provide a better description of pair production in this setup than either the low or intermediate temperature regimes, which would imply that the diagrams of Fig.~\ref{fig:feyn} do not dominate pair production and there is a nonperturbative enhancement over both pure Schwinger and pure thermal pair production.

Physically, it is clear that the process of pair production should not depend on the photon gas being precisely in equilibrium: if we use the picture of tunnelling from an excited state, one would expect that it is the energy and density of the photon distribution, rather than the nearness to equilibrium, that matters.

On the other hand, the condition of equilibrium is necessary for the calculation because it leads to important simplifications in the calculation of the production rate. The nonperturbative calculation of Eq.~\eqref{eq:rate_sphaleron} \cite{gould2017thermal,Gould:2018ovk} relied heavily on the Matsubara formalism~\cite{Bloch1932,Matsubara:1955ws}, which is only valid in equilibrium. In the high temperature regime a resummation of all-orders of the perturbative loop expansion was required. Generalising the result to any out-of-equilibrium distribution is beyond the scope of this paper.

We note however that the diagrams of Fig.~\ref{fig:feyn} can be calculated in an arbitrary photon distribution, following the approach of Ref.~\cite{Torgrimsson:2019sjn}, though in the high temperature regime these diagrams are not dominant. Considering the calculation in this distribution, it can be seen that these diagrams reproduce the perturbative Breit-Wheeler rate up to very small corrections, essentially because the photon gas is highly occupied at energies much greater than $k_B T_{CW}$ (see Eqs.~\eqref{eq:lowT} and \eqref{eq:tcw}). Further, perturbative corrections in this distribution due to the high intensity laser require one photon from the high energy tail of the distribution and hence are suppressed by $\exp(-\lambda m^2 c^3/(2\pi \hbar k_B T) + \mu/k_B T)\sim 10^{-10^{5}}$, where $T$ and $\mu$ here refer to the green dot-dashed line in Fig.~\ref{fig:intensity}. Thus any enhancement due to the high intensity laser must be a nonperturbative phenomenon which goes beyond the diagrams of Fig.~\ref{fig:feyn}.

Because the full nonperturbative calculation of the rate of pair production is beyond the scope of this paper, the possibility of nonperturbative enhancements in our proposed setup is conjectural. However, as all the effective temperatures we have used to describe the photon distribution are larger than $T_{WS}$, Eq.~\eqref{eq:tws}, we expect the high temperature regime to best describe the photon distribution in question. We thus expect a nonperturbative enhancement over the perturbative prediction, as is the case in equilibrium where the enhancement to the positron yield was $O(10^6)$. The experiment we have proposed here would be able to test this plausible conjecture, by performing null shots without the counterpropagating laser beams, for which only the perturbative process is possible. This would be able to determine which features of a photon distribution are important for the nonperturbative enhancements to pair production which feature in the thermal Schwinger effect, and how generic such enhancements are.

\section*{Acknowledgements}
O.G. would like to thank Holger Gies and Greger Torgrimsson for discussions related to this work. O.G. was supported from the Research Funds of the University of Helsinki. S.M. was supported by Engineering and Physical Sciences Research Council grant No. EP/M018555/1 and by Horizon 2020 under European Research Council Grant Agreement No. 682399. A.R. was supported by the U.K. Science and Technology Facilities Council grant ST/P000762/1. 

\bibliography{refs}

\end{document}